\documentstyle[12pt]{article}
\begin{document}
\centerline{\large\bf Effects of the $K^+\to\pi^+\nu\bar{\nu}$ and of other 
processes}
\vskip 0.6truecm
\centerline{\large\bf on the mixing hierarchies in the four-generation model}
\baselineskip=8truemm
\vskip 2.5truecm
\centerline{Toshihiko Hattori$,^{a),}$\footnote{e-mail: 
hattori@ias.tokushima-u.ac.jp} \ Tsutom Hasuike$,^{b),}$\footnote{e-mail: 
hasuike@anan-nct.ac.jp} \ and \ Seiichi Wakaizumi$ ^{c),}$\footnote{e-mail: 
wakaizum@medsci.tokushima-u.ac.jp}}
\vskip 0.6truecm
\centerline{\it $ ^{a)}$Institute of Theoretical Physics, University of Tokushima,
Tokushima 770-8502, Japan}
\centerline{\it $ ^{b)}$Department of Physics, Anan College of Technology,
Anan 774-0017, Japan}
\centerline{\it $ ^{c)}$School of Medical Sciences, University of Tokushima,
Tokushima 770-8509, Japan}
\vskip 2.5truecm
\centerline{\bf Abstract}
\vskip 0.7truecm

We analyze in the four-generation model the first measurement of the branching 
ratio of rare kaon decay $K^+\to\pi^+\nu\bar{\nu}$, along with the other 
processes of $K_L-K_S$ mass difference $\Delta m_K$, CP-violating 
parameter $\varepsilon_K, B_d-\bar{B_d}$ mixing, 
$B_s-\bar{B_s}$ mixing, $B(K_L\to\mu\bar{\mu})$, and the upper bound 
values of $D^0-\bar{D^0}$ mixing and $B(K_L\to\pi^0\nu\bar{\nu})$, and 
try to search for mixing of the fourth generation in the hierarchical mixing 
scheme of the Wolfenstein parametrization. 
Using the results for the mixing of the fourth generation, we discuss predictions 
of the $D^0-\bar{D^0}$ mixing$(\Delta m_D)$ and the branching ratio of 
directly CP-violating decay process $K_L\to\pi^0\nu\bar{\nu}$, and the 
effects on the CP asymmetry in neutral B meson decays and the unitarity triangle.
\newpage

\centerline{\large\bf I  Introduction}
\vskip 0.2truecm
In the physics of quarks and leptons, it has been long since the Standard Model 
acquired a remarkable success. As is shown, however, in the issue of mass 
generation of quarks and leptons and the physics such as SUSY, physics beyond 
the Standard Model has become highly expected. In this direction, 
the flavor-changing neutral current(FCNC) processes play important roles through 
the one-loop effects for the search of additional Higgs, new gauge bosons, 
additional fermions, etc.

Here we focus on the new branching ratio of the FCNC process $K^+\to\pi^+
\nu\bar{\nu}$, which is measured for the first time at the Brookhaven National 
Laboratory, $B=(4.2^{+9.7}_{-3.5})\times 10^{-10}$\cite{Adler}. It should 
be remarked that the central value is 4-6 times larger than 
the Standard Model prediction, $B=(0.6-1.5)\times 10^{-10}$\cite{Buchalla}, 
though the measurement is consistent with the theory within the experimental 
errors.

This process, $K^+\to\pi^+\nu\bar{\nu}$, had already been studied by 
Gaillard and Lee in 1974 and they obtained a branching ratio of $\sim 10^{-10}$ 
by using the short-distance 
$W-W$ box and $Z^0$-penguin diagrams in the "4-quark" model\cite{Gaillard}.
After that in 1981, Inami and Lim derived the rigorous expressions for these 
and other related diagrams relevant to the FCNC processes and studied the effects 
of superheavy quarks and leptons in $K_L\to \mu\bar{\mu}, K^+\to\pi^+\nu
\bar{\nu}$ and $K^0-\bar{K^0}$ mixing\cite{Inami}, before the top-quark 
was discovered.

In this work, we analyze the new branching ratio of $K^+\to\pi^+\nu
\bar{\nu}$ in the four-generation model\cite{Hayashi}\cite{Bigi}\cite{Hasuike}
under the expectation that the above-mentioned factor 4-6 of the measured value 
relative to the Standard Model predictions may imply the existence of a fourth 
generation with roughly the same mixing as for the third generation. 
We will investigate various possible mixings  for the fourth generation 
by imposing the constraints of $K^+\to\pi^+\nu\bar{\nu}$ and other processes 
of $K_L-K_S$ mass difference $\Delta m_K$, CP-violating parameter 
$\varepsilon_K, B_d-\bar{B_d}$ mixing, 
$B_s-\bar{B_s}$ mixing, $D^0-\bar{D^0}$ 
mixing, $B(K_L\to\pi^0\nu\bar{\nu})$ and $B(K_L\to\mu\bar{\mu})$, 
and we will study its effects on the $D^0-\bar{D^0}$ mixing and $B(K_L\to
\pi^0\nu\bar{\nu})$, of which only the upper bounds are experimentally 
known, CP violation in neutral $B$ meson decays and the unitarity triangle.

The paper is organized as follows. The four-generation model we use here is 
presented in Sec. II. In Sec. III we describe the phenomenological constraints 
on the model to search for possible mixings of the fourth generation. In Sec. IV 
we derive the "maximum" mixings allowed by the constraints. In Sec. V we 
discuss the consequences of the mixings on the $D^0-
\bar{D^0}$ mixing and the branching ratio of another FCNC decay process 
$B(K_L\to\pi^0\nu\bar{\nu})$, CP asymmetry in $B_d$ meson decays 
and the unitarity triangle, and finally we give conclusions. 
\vskip 0.6truecm

\centerline{\large\bf II  The four-generation model}
\vskip 0.2truecm

For the unitary $4\times 4$ quark mixing matrix in the four-generation scheme, 
we will use the Hou-Soni-Steger parametrization\cite{Hou}. The form of this 
parametrization is so complicated that we will not cite it here. It has, however, 
an advantage over the others that the third column and the fourth row have 
simple forms such that $(V_{ub},V_{cb},V_{tb},V_{t'b})=(s_zc_u
{\rm e}^{-{\rm i}\phi_1}, s_yc_zc_u, c_yc_zc_u, -s_u)$ and $(V_{t'd},
V_{t's},V_{t'b},V_{t'b'})=(-c_uc_vs_w{\rm e}^{{\rm i}\phi_3}, -c_us_v
{\rm e}^{{\rm i}\phi_2}, -s_u, c_uc_vc_w)$, and $V_{us}=s_xc_zc_v
-s_zs_us_v{\rm e}^{{\rm i}(\phi_2-\phi_1)}$, so that the three mixing 
angles $s_x(\equiv\sin\theta_x), s_y$ and $s_z$ give the elements $V_{us}, 
V_{cb}$ and $V_{ub}$ respectively as in the Standard Model, and the 
phase $\phi_1$ corresponds to the Kobayashi-Maskawa(KM) CP-violating 
phase $\delta^{KM}$\cite{Kobayashi}. The angles $s_u(\equiv\sin\theta_u), 
s_v$ and $s_w$, which give the elements $V_{t'b}, V_{t's}$ and $V_{t'd}$ 
respectively, are new mixing angles, and $\phi_2$ and $\phi_3$ are 
new phases. $t'$ and $b'$ are the fourth generation up- and down-quark, 
respectively. 

Since the magnitude of the three elements $V_{us}, V_{cb}$ and $V_{ub}$ 
are experimentally determined from the semileptonic decays of hyperons, $B$ 
mesons to hadrons with $c$- and $u$-quark, respectively, and are not 
affected by the existence of the fourth generation, we use the same values 
for the three angles $s_x, s_y$ and $s_z$ as in the 
Standard Model\cite{Buchalla} as an input of our analysis; 
\begin{equation}
s_x=0.22, \qquad s_y=0.040\pm 0.003, \qquad s_z/s_y=0.08\pm 0.02,  
\label{shiki1}
\end{equation}

We search for the mixings of the fourth generation allowed by the experimental 
quantities related to various FCNC processes. The mixing among the three 
generations in the Standard Model is known to be hierarchical as is well 
expressed by the Wolfenstein parametrization\cite{Wolfenstein1},
\begin{eqnarray}
V^{(3)} &=& \pmatrix{
V_{ud} & V_{us} & V_{ub} \cr
V_{cd} & V_{cs} & V_{cb} \cr
V_{td} & V_{ts} & V_{tb} \cr 
} \nonumber   \\
              &\simeq& \pmatrix{
1-\lambda^2/2 & \lambda & A(\rho -{\rm i}\eta)\lambda^3 \cr
-\lambda & 1-\lambda^2/2 & A\lambda^2 \cr
A(1-\rho-{\rm i}\eta)\lambda^3 & -A\lambda^2 & 1 \cr
} ,  \label{shiki2}
\end{eqnarray}
where $\lambda \equiv \sin\theta_C(\simeq 0.22)$ is the expansion parameter 
in the Wolfenstein parametrization. In the spirit of this parametrization, we will 
study the following cases of the fourth generation mixing to derive a "maximum" 
one allowed by the above-mentioned constraints;
\begin{eqnarray}
(V_{t'd},V_{t's},V_{t'b},V_{t'b'}) \simeq&(\lambda^5, \lambda^4, 
\lambda^3, 1), \nonumber \\
&(\lambda^4, \lambda^3, \lambda^2, 1), \nonumber \\
&(\lambda^3, \lambda^2, \lambda, 1), \nonumber \\
&(\lambda^2, \lambda^2, \lambda, 1), \nonumber \\
&(\lambda^3, \lambda^2, 1, \lambda), \nonumber \\
&(\lambda^2, \lambda, 1, \lambda), \nonumber \\
&(0, \lambda^3, \lambda, 1), \nonumber \\
&(0, \lambda^2, \lambda, 1).  \label{shiki3}
\end{eqnarray}
Here, we are not interested in the last two cases with $V_{t'd}=0$, because we 
will focus on the factor 4-6 of the central value of the measured branching ratio 
of $K^+\to\pi^+\nu\bar{\nu}$, relative to the predicted value in the Standard 
Model. 

\begin{table}
\caption{Combinations of relevant mixing matrix elements for 
$\Delta m_{B_d}, b\to s\gamma, K^+\to\pi^+\nu\bar{\nu}$ and 
$(K_L\to\mu\bar{\mu})_{\rm SD}$ for the third generation in the Standard 
Model and the four cases of the fourth generation mixing.}
\begin{center}
\begin{tabular}{ccccc}  \hline\hline
Mixing & $\Delta m_{B_d}$ & $b\to s\gamma$ & $K^+\to\pi^+\nu
\bar{\nu}$ & $(K_L\to\mu\bar{\mu})_{\rm SD}$ \\  \hline
$(V_{td},V_{ts},V_{tb})$ & $V_{td}V_{tb}$ & $V_{ts}V_{tb}$ & 
$V_{td}V_{ts}$ & $V_{td}V_{ts}$ \\
$(\lambda^3,\lambda^2,1)$ & $\lambda^3$ & $\lambda^2$ & $\lambda^5$ 
& $\lambda^5$ \\  \hline
$(V_{t'd},V_{t's},V_{t'b})$ & $V_{t'd}V_{t'b}$ & $V_{t's}V_{t'b}$ & 
$V_{t'd}V_{t's}$ & $V_{t'd}V_{t's}$ \\
$(\lambda^5,\lambda^4,\lambda^3)$ & $\lambda^8$ & $\lambda^7$ & 
$\lambda^9$ & $\lambda^9$ \\
$(\lambda^4,\lambda^3,\lambda^2)$ & $\lambda^6$ & $\lambda^5$ & 
$\lambda^7$ & $\lambda^7$ \\
$(\lambda^3,\lambda^2,\lambda)$ & $\lambda^4$ & $\lambda^3$ & 
$\lambda^5$ & $\lambda^5$ \\
$(\lambda^2,\lambda^2,\lambda)$ & $\lambda^3$ & $\lambda^3$ & 
$\lambda^4$ & $\lambda^4$ \\  \hline\hline
\end{tabular}
\end{center}
\label{tab1}
\end{table}
\vskip 0.1truecm

Table 1 shows the products of the relevant mixing matrix elements of the 
dominant contributions to the one-loop diagrams in $B_d-\bar{B_d}$ mixing
$(\Delta m_{B_d})$, $b\to s\gamma$ decay, $K^+\to\pi^+\nu\bar{\nu}$ 
decay and short-distance contributions to $K_L\to\mu\bar{\mu} ((K_L\to
\mu\bar{\mu})_{\rm SD})$ for the Standard Model and the four-generation 
model with the first four cases of mixing of eq.(\ref{shiki3}). As seen in 
Table 1, the first two 
cases of $(V_{t'd},V_{t's},V_{t'b}) \simeq (\lambda^5, \lambda^4, 
\lambda^3)$ and $(\lambda^4, \lambda^3, \lambda^2)$ give too small 
contributions to affect the branching ratio of $K^+\to\pi^+
\nu\bar{\nu}$ and they also do not give any significant 
contributions to $\Delta m_{B_d}$ and $(K_L\to\mu\bar{\mu})_{\rm SD}$. 
The third case of $(V_{t'd},V_{t's},V_{t'b}) \simeq (\lambda^3, \lambda^2, 
\lambda)$ gives the same order of contributions to $K^+\to\pi^+
\nu\bar{\nu}$ and $(K_L\to\mu\bar{\mu})_{\rm SD}$ as in the Standard 
Model. 
It turns out that even this favorable case of $(\lambda^3, \lambda^2, \lambda)$ 
does not contribute to $b\to s\gamma$ so much as in the Standard Model, so 
we will not include the process $b\to s\gamma$ in the following numerical 
analysis. Although the fifth and sixth cases of $(V_{t'd},V_{t's},V_{t'b},
V_{t'b'}) \simeq (\lambda^3, \lambda^2, 1, \lambda)$ and $(\lambda^2, 
\lambda, 1, \lambda)$ of eq.(\ref{shiki3}) are interesting, these cases have
 proved not to lead to any favorable solutions in our numerical analysis. 
\vskip 0.7truecm

\centerline{\large\bf III  Constraints on the model}
\vskip 0.2truecm

The constraints we impose on the model to search for the fourth generation 
mixing are the following, $K_L-K_S$ mass difference $\Delta m_K=(3.522
\pm 0.016)\times 10^{-12}$ MeV\cite{Particle}, CP-violating parameter 
in the neutral kaon system $\varepsilon_K=(2.28\pm 0.02)\times 10^{-3}$
\cite{Particle}, $\Delta m_{B_d}=(3.12\pm 0.20)\times 10^{-10}$ MeV
\cite{Particle} for $B_d-\bar{B_d}$ mixing, 
$B(K^+\to\pi^+\nu\bar{\nu})=(4.2^{+9.7}_{-3.5})\times 10^{-10}
$\cite{Adler}, $\Delta m_{B_s}>52.0\times 10^{-10}$ MeV\cite{Adam} 
for $B_s-\bar{B_s}$ mixing, 
$\Delta m_D<1.4\times 10^{-10}$ MeV\cite{Aitala} for $D^0-\bar{D^0}$ 
mixing, $B(K_L\to\pi^0\nu\bar{\nu})<5.8\times 10^{-5}$\cite{Weaver} and 
$B(K_L\to\mu\bar{\mu})_{\rm SD}<2.2\times 10^{-9}$, where the upper 
bound of the short-distance(SD) contribution to $B(K_L\to\mu\bar{\mu})$ is 
taken to be the value estimated by B\'{e}langer and Geng\cite{Belanger}. 
As for the directly CP-violating parameter in the neutral kaon system 
$\varepsilon'/\varepsilon$, the experimental values by the two groups at 
CERN and Fermilab deviated from each other by more than 2.4 standard 
deviations and recently KTeV at Fermilab has obtained a completely consistent
value of ${\rm Re}(\varepsilon'/\varepsilon)=(28.0\pm 4.1)\times 10^{-4}$
\cite{KTeV} with the one by NA31 of ${\rm Re}(\varepsilon'/\varepsilon)=
(23\pm 7)\times 10^{-4}$\cite{epsilonprime}. The formulation of 
$\varepsilon'/\varepsilon$ in the four-generation model with appropriate 
QCD corrections is complicated and is out of scope in our paper\cite{Buras99}. 
So, we will not include $\varepsilon'/\varepsilon$ here. 

Each of the above-mentioned eight constraints is described in the following.
\vskip 0.3truecm

\noindent
(i)$K_L-K_S$ mass difference, $\Delta m_K$

\noindent
The short-distance part of $\Delta m_K$ comes from the well-known $W-W$ 
box diagram with $c, t$ and $t'$ as internal quarks as shown in Fig.1 in the 
four-generation model and the contribution is expressed, 
for example, for the box diagram with two $c$-quarks as follows,
\begin{equation}
\Delta m_K(c,c)=\frac{G_F^2M_W^2}{6\pi^2}f_K^2B_Km_K{\rm Re}
[(V_{cs}V_{cd}^*)^2]\eta^K_{cc}S(x_c),  \label{shiki4}
\end{equation}
where $S(x)$ is the Inami-Lim box function\cite{Inami}, $x_c\equiv m_c^2/
M_W^2$, $m_c$ being the charm-quark mass, $\eta^K_{cc}$ is the QCD 
correction factor including the next-to-leading order effects, and $f_K$ and $B_K$ 
are the decay constant and the bag parameter of the kaon, respectively. By taking 
for these parameters the values of $m_c=1.3$ GeV, $\eta^K_{cc}=1.38$
\cite{Buchalla}, $f_K=0.16$ GeV and $B_K=0.75\pm 0.15$\cite{Buchalla}, 
we obtain from the inputs of eq.(\ref{shiki1}) the $(c,c)$ contribution as $\Delta 
m_K(c,c)=(2.6-3.9)\times 10^{-12}$ MeV, which is already consistent by itself 
with the measured value. Numerically, $(c,t)$ and $(t,t)$ contributions are 
very small in comparison with the $(c,c)$ contribution, so we take a constraint for 
the fourth-generation contributions to be 
\begin{equation}
\left| \frac{\Delta m_K(c,t')+\Delta m_K(t,t')+\Delta m_K(t',t')}
{\Delta m_K(c,c)}\right| <1      \label{shiki5}
\end{equation}
as a loose constraint, since there are a large amount of the long-distance 
contributions. In eq.(\ref{shiki5}), $(t',t')$ contribution, $\Delta m_K(t',t')$, is 
given as follows, 
\begin{equation}
\Delta m_K(t',t')=\frac{G_F^2M_W^2}{6\pi^2}f_K^2B_Km_K{\rm Re}
[(V_{t's}V_{t'd}^*)^2]\eta^K_{t't'}S(x_{t'}),  \label{shiki6}
\end{equation}
where $x_{t'}\equiv m_{t'}^2/M_W^2$, $m_{t'}$ being the fourth-generation 
$t'$ mass, and $S(x_{t'})$ can be approximated as $0.707x_{t'}^{0.82}$ 
for $130\le m_{t'}\le 1200$GeV. $\eta^K_{t't'}$ is the QCD correction factor 
which is taken here to the leading order as 
\begin{equation}
\eta^K_{t't'} = [\alpha_s(m_c)]^{6/27}\left[ \frac{\alpha_s(m_b)}
{\alpha_s(m_c)} \right]^{6/25}\left[ \frac{\alpha_s(m_t)}{\alpha_s(m_b)} 
\right]^{6/23}\left[ \frac{\alpha_s(m_{b'})}{\alpha_s(m_t)} \right]^{6/21} 
\left[ \frac{\alpha_s(\mu_{t'})}{\alpha_s(m_{b'})} \right]^{6/19}.  
\label{shiki7}
\end{equation}
In eq.(\ref{shiki7}), $\alpha_s(m)$ is the running coupling constant in QCD and 
is expressed as 
\begin{equation}
\alpha_s(m)=\frac{4\pi}{\beta_0\ln(m^2/\Lambda^2)} ,   \label{shiki8}
\end{equation}
where $\Lambda$ is the QCD scale of 0.10GeV and $\beta_0=11-\frac{2}{3}
N_f$, $N_f$ being the number of active quark flavors at the relevant energy scale, 
and $\mu_{t'}\simeq O(m_{t'})$. $\eta^K_{t't'}$ turns out to be 0.61 for 
$m_c=1.3$GeV, $m_b=4.4$GeV, $m_t=180$GeV, $m_{b'}=370$GeV and 
$m_{t'}=400$GeV, the constraint on the fourth-generation quark masses being 
described at the end of this section. Similarly, $\Delta m_K(t,t')$ and 
$\Delta m_K(c,t')$ are expressed as 
\begin{eqnarray}
\Delta m_K(t,t')&=&2\frac{G_F^2M_W^2}{6\pi^2}f_K^2B_Km_K{\rm Re}
[V_{ts}V_{td}^*V_{t's}V_{t'd}^*]\eta^K_{tt'}S(x_t,x_{t'}),  \\ 
\label{shiki9}
\Delta m_K(c,t')&=&2\frac{G_F^2M_W^2}{6\pi^2}f_K^2B_Km_K{\rm Re}
[V_{cs}V_{cd}^*V_{t's}V_{t'd}^*]\eta^K_{ct'}S(x_c,x_{t'}),  
\label{shiki10}
\end{eqnarray}
where $S(x_t,x_{t'})$ is the Inami-Lim function for $W-W$ box diagram with $t$- 
and $t'$-quark in the internal line\cite{Inami} and the QCD correction factors 
$\eta^K_{tt'}$ and $\eta^K_{ct'}$ are taken as 0.5 and 0.6, respectively. 
\vskip 0.3truecm

\noindent
(ii)CP-violating parameter in neutral kaon system, $\varepsilon_K$

\noindent
The quantity $\varepsilon_K$ is expressed by the imaginary part of hadronic matrix 
element of the effective Hamiltonian with $\Delta S=2$ between $K^0$ and 
$\bar{K^0}$, to which the short-distance contribution comes from the $W-W$ 
box diagram as in $\Delta m_K$. The box contributions with $c$- and $t$-quark 
and with two $t$ quarks give the expressions of
\begin{eqnarray}
\varepsilon_K(c,t)&=&\frac{1}{\sqrt{2}\Delta m_K}
\frac{G_F^2M_W^2}{6\pi^2}f_K^2B_Km_K{\rm Im}[V_{cs}V_{cd}^*
V_{ts}V_{td}^*]\eta^K_{ct}S(x_c,x_t), \\     \label{shiki11}
\varepsilon_K(t,t)&=&\frac{1}{\sqrt{2}\Delta m_K}
\frac{G_F^2M_W^2}{12\pi^2}f_K^2B_Km_K{\rm Im}[(V_{ts}V_{td}^*)^2]
\eta^K_{tt}S(x_t).       \label{shiki12}
\end{eqnarray}
If we take the QCD correction factors including the next-to-leading order 
as $\eta_{ct}^K=$0.47 and $\eta_{tt}^K=$0.57\cite{Buchalla}, the dominant 
terms in the $(c,t)$- and $(t,t)$-box contributions lead to $\varepsilon_K(c,t)
\simeq 2.83\times 10^{-3}B_K\sin\phi_1$ and $\varepsilon_K(t,t)\simeq 
2.41\times 10^{-3}B_K\sin(2\phi_1)$ in the Standard Model, 
where $\phi_1$ is the CP-violating phase $\delta^{KM}$. Since the 
magnitude of these two contributions is already close to the measured value 
$\varepsilon_K=(2.28\pm 0.02)\times 10^{-3}$ by taking into consideration 
the theoretical uncertainty in the bag parameter value, $B_K=0.75\pm 0.15$, 
we take the constraint from $\varepsilon_K$ on the model that the sum of the 
contributions from $c, t$ and $t'$ quarks should be within the $1\sigma$ error 
of the measured value,
\begin{equation}
\sum_{i,j=c,t,t',i\le j}\varepsilon_K(i,j)=(2.28\pm 0.02)\times 10^{-3}.
\label{shiki13}
\end{equation}
\vskip 0.3truecm

\noindent
(iii)$B_d-\bar{B_d}$ mixing, $\Delta m_{B_d}$

\noindent
The mass difference between the two mass-eigenstates of $B_d-\bar{B_d}$ 
system is given by the $W-W$ box diagram, and the $(t,t)$-box contribution 
is expressed by
\begin{equation}
\Delta m_{B_d}(t,t)=\frac{G_F^2M_W^2}{6\pi^2}f_B^2B_Bm_{B_d}
\left| V_{tb}V_{td}^*\right| ^2\eta^B_{tt}S(x_t),   \label{shiki14}
\end{equation}
where $f_B$ and $B_B$ are the decay constant and the bag parameter for 
$B_d$ meson, respectively, and $\eta^B_{tt}$ is the QCD correction factor 
including the next-to-leading order effects. By taking for these parameters the 
values of $\sqrt{B_B}f_B=(0.20\pm 0.04)$ GeV and 
$\eta^B_{tt}=0.55$\cite{Buchalla} and by using the inputs of eq.(\ref{shiki1}), 
we obtain the $(t,t)$ contribution $\Delta m_{B_d}(t,t)=(1.75-3.95)\times 
10^{-10}$ MeV in the Standard Model. This value is consistent with the measured 
value, $\Delta m_{B_d}=(3.12\pm 0.20)\times 10^{-10}$ MeV\cite{Particle}. 
Since $(c,c)$ and $(c,t)$ contributions are numerically very small in comparison 
with the $(t,t)$ contribution, we take the constraint from $\Delta m_{B_d}$ on 
the model that the sum of the contributions from $t$ and $t'$ should be within the 
$1\sigma$ error of the measured value as follows, 
\begin{eqnarray}
\frac{G_F^2M_W^2}{6\pi^2}f_B^2B_Bm_{B_d}
&\times&\left| (V_{tb}V_{td}^*)^2\eta^B_{tt}S(x_t)+(V_{t'b}V_{t'd}^*)^2
\eta^B_{t't'}S(x_{t'})+2V_{tb}V_{td}^*V_{t'b}V_{t'd}^*\eta^B_{tt'}
S(x_t,x_{t'}) \right|    \nonumber \\
&=&(3.12\pm 0.20)\times 10^{-10} {\rm MeV},  \label{shiki15}
\end{eqnarray}
where we take for the QCD correction factor $\eta^B_{t't'}$ the following 
expression to the leading order, 
\begin{equation}
\eta^B_{t't'}=[\alpha_s(m_t)]^{6/23}\left[ \frac{\alpha_s(m_{b'})}
{\alpha_s(m_t)} \right]^{6/21}\left[ \frac{\alpha_s(\mu_{t'})}
{\alpha_s(m_{b'})} \right]^{6/19},  \label{shiki16}
\end{equation}
which turns out to be 0.58 for the same set of parameter values as for 
$\eta^K_{t't'}$. Another QCD correction factor $\eta^B_{tt'}$ in 
eq.(\ref{shiki15}) is taken as 0.5. 
\vskip 0.3truecm

\noindent
(iv) $B(K^+\to \pi^+\nu\bar{\nu})$

\noindent
The short-distance contributions to the FCNC decay $K^+\to\pi^+\nu\bar{\nu}$ 
come from the $W-W$ box diagram and $Z^0$-penguin diagrams as shown 
in Fig.2 in the four-generation model. The expression for the contributions 
including the next-to-leading order QCD effects is given by Buchalla and Buras
\cite{Buras93}\cite{Buras} in the Standard Model and are summarized in ref.2. 
We add to their expression of the branching ratio the contribution from $t'$-quark 
exchange as follows, 
\begin{equation}
B(K^+\to\pi^+\nu\bar{\nu})=\kappa_+\left| \frac{V_{cd}V_{cs}^*}{\lambda}
P_0+\frac{V_{td}V_{ts}^*}{\lambda^5}\eta_tX_0(x_t)+\frac{V_{t'd}
V_{t's}^*}{\lambda^5}\eta_{t'}X_0(x_{t'})\right|^2,      \label{shiki17}
\end{equation}
where $\kappa_+=4.57\times 10^{-11}$, $P_0$ is the sum
of charm contributions to the two diagrams including the next-to-leading order QCD 
corrections\cite{Buras} and $X_0$ is the sum of the $W-W$ box and 
$Z^0$-penguin functions without QCD corrections calculated by Inami and Lim
\cite{Inami}, the expressions of $P_0$ and $X_0$ being summarized in ref.2. In 
eq.(\ref{shiki17}), $\eta_t(=0.985)$ is the next-to-leading order QCD correction 
factor to the $t$-quark exchange\cite{Buchalla}\cite{Buras93}, and we take 
$\eta_{t'}=1.0$ for $t'$-exchange, since $\eta_t$ is almost 1.0 and the running 
distance for for the QCD 
corrections for $t'$-exchange is shorter for $m_{t'}>m_t$ than that for the 
$t$-exchange. The constraint is that the branching ratio of 
eq.(\ref{shiki17}) should be consistent with the measured value of branching 
ratio $B=(4.2^{+9.7}_{-3.5})\times 10^{-10}$\cite{Adler}, since the 
long-distance contribution is estimated to be very small $(B\sim 10^{-13})$
\cite{Rein}. We do not take into consideration the mixing effect in the leptonic 
sector.
\vskip 0.3truecm

\noindent
(v) $B_s-\bar{B_s}$ mixing

\noindent
The dominant contributions to $B_s-\bar{B_s}$ mixing are the $W-W$ box 
diagrams with $t$ and $t'$ exchanges as in $B_d-\bar{B_d}$ mixing. We take 
the constraint that the sum of $(t, t), (t, t')$ and $(t', t')$ contributions to 
$\Delta m_{B_s}$ should be larger than the present experimental lower bound; 
$\Delta m_{B_s}>52.0\times 10^{-10}$ MeV \cite{Adam}, where 
$\Delta m_{B_s}$ is the mass difference of the two mass 
eigenstates of $B_s-\bar{B_s}$ system. The constraint is expressed as follows, 
\begin{eqnarray}
\frac{G_F^2M_W^2}{6\pi^2}f_{B_s}^2B_{B_s}m_{B_s}
&\times&\left| (V_{tb}V_{ts}^*)^2\eta^B_{tt}S(x_t)+(V_{t'b}V_{t's}^*)^2
\eta^B_{t't'}S(x_{t'})+2V_{tb}V_{ts}^*V_{t'b}V_{t's}^*\eta^B_{tt'}
S(x_t,x_{t'}) \right|   \nonumber  \\
&>& 52.0\times 10^{-10} \quad {\rm MeV}.  \label{shiki18}
\end{eqnarray}
We take the quantity $\sqrt{B_{B_s}}f_{B_s}$ to be equal to that for 
$B_d-\bar{B_d}$ mixing, and the QCD correction factors $\eta^B_{tt}, 
\eta^B_{t't'}$ and $\eta^B_{tt'}$ are equal to the ones for $B_d-\bar{B_d}$ 
mixing.
\vskip 0.3truecm

\noindent
(vi) $D^0-\bar{D^0}$ mixing

\noindent
The dominant contribution to $D^0-\bar{D^0}$ mixing in the four-generation 
model is the $W-W$ box diagram with fourth-generation down-quark $b'$ 
exchange\cite{Babu} as shown in Fig.3. We take the constraint that this 
contribution to the mass difference between the two mass-eigenstates of 
$D^0-\bar{D^0}$ system should be smaller than the present experimental 
upper bound\cite{Aitala}, $\Delta m_D(b',b')<1.4\times 10^{-10} 
{\rm MeV}$, since the Standard Model box contribution of two $s$-quarks 
exchange\cite{Datta} and the long-distance contributions\cite{Wolfenstein} are 
estimated to be three to four orders of magnitude smaller than the upper bound. 
The constraint is expressed as follows,
\begin{equation}
\Delta m_D(b',b')=\frac{G_F^2M_W^2}{6\pi^2}f_D^2B_Dm_D{\rm Re}
[(V_{cb'}^*V_{ub'})^2]\eta^D_{b'b'}S(x_{b'})<1.4\times 10^{-10} 
\quad {\rm MeV},  \label{shiki19}
\end{equation}
where $x_{b'}\equiv m_{b'}^2/M_W^2$. We take for the QCD correction 
factor $\eta^D_{b'b'}$ the following expression to the leading order, 
\begin{equation}
\eta^D_{b'b'}=[\alpha_s(m_b)]^{6/25}\left[ \frac{\alpha_s(m_t)}
{\alpha_s(m_b)} \right]^{6/23}\left[ \frac{\alpha_s(\mu_{b'})}
{\alpha_s(m_t)} \right]^{6/21},  \label{shiki20}
\end{equation}
which is about 0.58 for $\mu_{b'}\simeq m_{b'}=370$ GeV, $m_t=180$ GeV 
and $m_b=4.4$ GeV. We tentatively take $f_D\sqrt{B_D}=$0.2 GeV in the 
following numerical analyses, since the numerical result of $\Delta m_D(b',b')$ is 
of the order of $10^{-12}$ MeV for the range of $f_D\sqrt{B_D}=(0.1-0.3)$GeV. 
Incidentally, the Standard Model prediction of $\Delta m_D$ is around 
$10^{-14}$MeV\cite{Babu}.
\vskip 0.3truecm

\noindent
(vii) $B(K_L\to \pi^0\nu\bar{\nu})$

\noindent
The process $K_L\to \pi^0\nu\bar{\nu}$ is the "direct" CP-violating decay
\cite{Littenberg} and the rate is expressed by the imaginary part of sum of the 
same $W-W$ box and $Z^0$-penguin diagram amplitudes as in $K^+\to \pi^+
\nu\bar{\nu}$\cite{Buchalla}, since the CP-conserving contribution is known 
to be very strongly suppressed\cite{Isidori}. Therefore, we take the constraint 
that the sum of $t$ and $t'$ contributions to the branching ratio should be smaller 
than the experimental upper bound\cite{Weaver}, $B(K_L\to\pi^0\nu\bar{\nu})
<5.8\times 10^{-5}$. The constraint is expressed as 
\begin{equation}
\kappa_L\left( \frac{{\rm Im}(V_{td}V_{ts}^*)}{\lambda^5}\eta_tX_0(x_t)
+\frac{{\rm Im}(V_{t'd}V_{t's}^*)}{\lambda^5}\eta_{t'}X_0(x_{t'}) 
\right)^2<5.8\times 10^{-5} ,      \label{shiki21}
\end{equation}
where $\kappa_L=1.91\times 10^{-10}$, $X_0$ is the same function and 
$\eta_t$ and $\eta_{t'}$ are the same QCD correction factors as appeared in 
eq.(\ref{shiki17}) for $K^+\to\pi^+\nu\bar{\nu}$. 
\vskip 0.3truecm

\noindent
(viii) $B(K_L\to\mu\bar{\mu})_{\rm SD}$

\noindent
The process $K_L\to\mu\bar{\mu}$ is the CP-conserving decay. The 
short-distance(SD) contribution is given by the $W-W$ box and 
$Z^0$-penguin diagrams and the branching ratio for this part is expressed as
\cite{Buchalla}
\begin{equation}
B(K_L\to\mu\bar{\mu})_{\rm SD}=\kappa_{\mu} \left[ \frac{{\rm Re}
\left( V_{cd}V_{cs}^*\right) }{\lambda}P'_0+\frac{{\rm Re}\left( V_{td} 
V_{ts}^*\right) }{\lambda^5}\eta^Y_tY_0(x_t)+\frac{{\rm Re}\left( V_{t'd}
V_{t's}^*\right) }{\lambda^5}\eta^Y_{t'}Y_0(x_{t'})\right]^2, 
   \label{shiki22}
\end{equation}
where $\kappa_{\mu}=1.68\times 10^{-9}$, $P'_0$ is the sum 
of charm contributions to the two diagrams including the next-to-leading order 
QCD corrections\cite{Buras} and $Y_0$ the sum 
of the $W-W$ box and $Z^0$-penguin functions without QCD corrections 
calculated by Inami and Lim\cite{Inami}, the expressions of $P'_0$ and $Y_0$ 
being summarized in ref.2. In eq.(\ref{shiki22}), $\eta^Y_t(=1.026)$ is the 
next-to-leading order QCD correction factor to the $t$-quark exchange
\cite{Buchalla}\cite{Buras93} and we take $\eta^Y_{t'}=1.0$ for $t'$ exchange 
for the same reason as stated for $K^+\to\pi^+\nu\bar{\nu}$. 
We take the constraint that the branching ratio of 
eq.(\ref{shiki22}) should be smaller than the upper bound of the short-distance 
contribution\cite{Belanger} as stated before at the beginning of this section, 
$B(K_L\to\mu\bar{\mu})_{\rm SD}<2.2\times 10^{-9}$. We do not take 
into consideration the mixing effect in the leptonic sector. 

For the masses of $t'$ and $b'$, there is a constraint from $\rho$ parameter. 
If we denote the parameter $\rho_0$ as 
\begin{equation}
\rho_0 = \frac{M_W^2}{M_Z^2\cos^2\theta_W\hat{\rho}} ,  \label{shiki23}
\end{equation}
where $\sin^2\theta_W$ is the Weinberg angle and $\hat{\rho}$ is the quantity 
$M_W^2/(M_Z^2\cos^2\theta_W)$, which involves the radiative correction 
effects from Higgs doublets and top-quark mass, then $\rho_0-1$ describes 
new sources of SU(2) breaking. The fourth generation makes $\rho_0$ deviate 
from 1 as\cite{PDG}
\begin{equation}
\rho_0=1+\frac{3G_F}{8{\sqrt 2}\pi^2}\left( m_{t'}^2+m_{b'}^2
-\frac{4m_{t'}^2m_{b'}^2}{m_{t'}^2-m_{b'}^2}\ln\frac{m_{t'}}{m_{b'}} 
\right) .  \label{shiki24}
\end{equation}
The value of $\rho_0$ is now $\rho_0=0.9998\pm0.0008$\cite{PDG}, and 
this constrains the masses of $t'$ and $b'$. 
\vskip 0.6truecm

\centerline{\large\bf IV  Possible mixings of fourth generation}
\vskip 0.3truecm

We search for possible mixings of the fourth generation allowed by the eight 
constraints in the previous section by testing the typical hierarchical mixings of 
eq.(\ref{shiki3}) with the intention to obtain the "maximum" mixing compatible 
with the considerably large branching ratio of the rare decay $K^+\to\pi^+
\nu\bar{\nu}$ with a factor of 4-6 as compared with the predictions in the 
Standard Model. From this point of view, the last two cases with $V_{t'd}=0$ 
of eq.(\ref{shiki3}) are not interesting here. 

Free parameters are the three phases $\phi_1, \phi_2$ and $\phi_3$ of the 
$4\times 4$ mixing matrix. As for the masses of the fourth 
generation quarks, we choose tentatively $(m_{t'}, m_{b'})=(400, 370), 
(800, 770)$ and $(1200, 1170)$ GeV as typical ones, which are compatible 
with the constraint of 
eq.(\ref{shiki24}). We vary the three phases in the range of 
$0\le\phi_1, \phi_2, \phi_3\le2\pi$. We found no solutions compatible with 
the eight constraints for the exotic fifth and sixth cases of $(V_{t'd},V_{t's},
V_{t'b},V_{t'b'}) \simeq (\lambda^3, \lambda^2, 1, \lambda)$ and 
$(\lambda^2, \lambda, 1, \lambda)$ of eq.(\ref{shiki3}). So, we focus on 
the first four cases of eq.(\ref{shiki3}) here. 

Strong constraints come from $\Delta m_K, \varepsilon_K, B_d-\bar{B_d}$ 
mixing, $K^+\to\pi^+\nu\bar{\nu}$ and $(K_L\to\mu
\bar{\mu})_{\rm SD}$. In the Standard Model, the largest contribution comes 
from the top-quarks for $B_d-\bar{B_d}$ mixing, $K^+\to
\pi^+\nu\bar{\nu}$ and $(K_L\to\mu\bar{\mu})_{\rm SD}$, and there 
the combination of the relevant CKM matrix elements is $V_{td}V_{tb}
\sim\lambda^3$ for $B_d-\bar{B_d}$ mixing and $V_{td}V_{ts}\sim
\lambda^5$ for $K^+\to\pi^+\nu\bar{\nu}$ and 
$(K_L\to\mu\bar{\mu})_{\rm SD}$. On the other hand, 
the combinations of the corresponding matrix elements for $t'$-quark are 
shown in Table 1 for each of the above four cases. By comparing these 
combinations between the Standard Model and the four-generation model, 
the numerical analyses give the following results; the cases of $(s_w,s_v,s_u)\ 
(\simeq (|V_{t'd}|,|V_{t's}|,|V_{t'b}|)) = (\lambda^5,\lambda^4,\lambda^3)$ 
and $(\lambda^4,\lambda^3,\lambda^2)$ give almost the same predictions to the 
above-mentioned eight quantities as in the Standard Model, since the contributions
of the fourth generation are very small, as seen from Table 1. 
For the case of $(\lambda^3,\lambda^2,\lambda)$, almost all the quantities 
satisfy the constraints with only one exception of $B(K_L\to\mu
\bar{\mu})_{\rm SD}$, for which this mixing gives a value several 
times larger than the upper bound. The last case of
$(\lambda^2,\lambda^2,\lambda)$ predicts too large values for $B(K^+\to
\pi^+\nu\bar{\nu})$ and $B(K_L\to\mu\bar{\mu})_{\rm SD}$.
These results imply that the mixing of $(\lambda^3,\lambda^2,\lambda)$ is 
a little large for the fourth generation and it turns out that a mixing with $s_w$ 
and $s_v$ reduced by $20\%$, that is, $(s_w,s_v,s_u)= (0.8\lambda^3,
0.8\lambda^2,\lambda)$ satisfies all of the eight constraints for 
$(m_{t'},m_{b'})=(400, 370)$ GeV, the one with $s_w$ and $s_v$ reduced 
by $50\%$, that is, $(s_w,s_v,s_u)= (0.5\lambda^3,0.5\lambda^2,\lambda)$ 
satisfies them for $(m_{t'},m_{b'})=(800, 770)$ GeV and the one with  
$(s_w,s_v,s_u)= (0.3\lambda^3,0.3\lambda^2,\lambda)$ does 
for $(m_{t'},m_{b'})=(1200, 1170)$ GeV as a maximum mixing. This strong 
energy-dependence of the reduction factors $(s_w/\lambda^3, s_v/\lambda^2, 
s_u/\lambda)$ is valid and reasonable, because the contribution of the $t'$-quark 
exchange to the decay amplitudes of both $K^+\to\pi^+\nu\bar{\nu}$ 
and $K_L\to\pi^0\nu\bar{\nu}$ is proportional to $V_{t'd}V_{t's}^*
X_0(x_{t'})\simeq 
\frac{1}{8} s_ws_v(m_{t'}/M_W)^2{\rm e}^{{\rm i}(\phi_3-\phi_2)}$, 
the one to the amplitude of $(K_L\to \mu\bar{\mu})_{\rm SD}$ is 
${\rm Re}(V_{t'd}V_{t's}^*)Y_0(x_{t'})\simeq \frac{1}{8}s_ws_v
(m_{t'}/M_W)^2\cos (\phi_3-\phi_2)$, the contribution to $\Delta m_K$ is 
${\rm Re}[(V_{t'd}^*V_{t's})^2]S(x_{t'})\simeq 0.707s_w^2s_v^2
(m_{t'}/M_W)^{1.64}\cos 2(\phi_2-\phi_3)$, and the one to 
$\Delta m_{B_d}$ is $|V_{t'd}^*V_{t'b}|^2S(x_{t'})\simeq 0.707
s_w^2s_u^2(m_{t'}/M_W)^{1.64}$. 

\begin{table}
\caption{Typical solutions for the fourth-generation mixing $(s_w,s_v,s_u)=
(0.8\lambda^3,0.8\lambda^2,\lambda)$ in case of $(m_{t'},m_{b'})=
(400, 370)$GeV for the three phases of $(\phi_1, \phi_2, \phi_3)$. Predictions 
of $B(K^+\to \pi^+\nu\bar{\nu})(10^{-10}), \Delta m_D({\rm in}  10^{-12}
{\rm MeV}), B(K_L\to \pi^0\nu\bar{\nu})(10^{-10})$ and the CP 
asymmetry for $B_d\to J/\psi K_S$ are added.}
\begin{center}
\begin{tabular}{c|c|c|c|c|c|c} \hline\hline
$\phi_1$ & $\phi_2$ & $\phi_3$ & $K^+\to \pi^+\nu\bar{\nu}$ & 
$\Delta m_D$ & $K_L\to \pi^0\nu\bar{\nu}$ & $C_f(B_d\to 
J/\psi K_S)$ \\ 
\hline
$3\pi/2$ & $\pi/6$ & $\pi/2$ & 2.8 & 0.8 & 11.7 & -0.37 \\
$\pi/4$ & $\pi/3$ & 0 & 2.2 & 1.8 & 8.7 & 0.24 \\
$3\pi/2$ & $\pi/2$ & $\pi/4$ & 2.3 & 1.6 & 8.6 & -0.34 \\
$7\pi/12$ & $\pi/2$ & $3\pi/4$ & 4.1 & 0.6 & 16.7 & 0.26 \\
$\pi/2$ & $5\pi/6$ & $\pi/2$ & 1.7 & 0.7 & 7.0 & 0.25 \\
$3\pi/4$ & $\pi$ & $3\pi/4$ & 1.5 & 0.8 & 5.8 & 0.17 \\
$13\pi/8$ & $\pi$ & $3\pi/4$ & 2.7 & 0.7 & 9.9 & -0.35 \\
$\pi/2$ & $7\pi/6$ & $11\pi/12$ & 1.7 & 0.4 & 6.6 & 0.30 \\
$\pi/3$ & $4\pi/3$ & $13\pi/12$ & 2.1 & 0.4 & 8.0 & 0.32 \\
$\pi/2$ & $2\pi$ & $7\pi/4$ & 1.6 & 2.0 & 6.2 & 0.34 \\ \hline
\end{tabular}
\end{center}
\label{tab2}
\end{table}
\vskip 0.1truecm

\begin{table}
\caption{Typical solutions for the fourth-generation mixing $(s_w,s_v,s_u)=
(0.5\lambda^3,0.5\lambda^2,\lambda)$ in case of $(m_{t'},m_{b'})=
(800, 770)$GeV for the three phases of $(\phi_1, \phi_2, \phi_3)$. Predictions 
of $B(K^+\to \pi^+\nu\bar{\nu})(10^{-10}), \Delta m_D({\rm in}  10^{-12}
{\rm MeV}), B(K_L\to \pi^0\nu\bar{\nu})(10^{-10})$ and the CP 
asymmetry for $B_d\to J/\psi K_S$ are added.}
\begin{center}
\begin{tabular}{c|c|c|c|c|c|c} \hline\hline
$\phi_1$ & $\phi_2$ & $\phi_3$ & $K^+\to \pi^+\nu\bar{\nu}$ & 
$\Delta m_D$ & $K_L\to \pi^0\nu\bar{\nu}$ & $C_f(B_d\to 
J/\psi K_S)$ \\ 
\hline
$5\pi/12$ & $\pi/3$ & 0 & 2.8 & 1.1 & 11.5 & 0.26 \\
$\pi/2$ & $\pi/2$ & $\pi/6$ & 2.6 & 0.9 & 10.6 & 0.19 \\
$5\pi/12$ & $2\pi/3$ & $\pi/3$ & 2.7 & 0.6 & 10.8 & 0.17 \\
$\pi/2$ & $5\pi/6$ & $\pi/2$ & 2.7 & 0.3 & 11.0 & 0.22 \\
$\pi/2$ & $\pi$ & $4\pi/3$ & 4.8 & 0.1 & 19.8 & 0.38 \\
$7\pi/12$ & $7\pi/6$ & $3\pi/2$ & 4.5 & 0.3 & 18.5 & 0.37 \\
$15\pi/8$ & $3\pi/2$ & 0 & 5.1 & 0.4 & 19.5 & -0.13 \\
$\pi/2$ & $3\pi/2$ & $11\pi/6$ & 4.6 & 0.7 & 19.2 & 0.32 \\
$7\pi/4$ & $5\pi/3$ & $\pi/6$ & 5.0 & 0.5 & 18.6 & -0.32 \\
$19\pi/12$ & $11\pi/6$ & $\pi/3$ & 5.2 & 0.4 & 18.0 & -0.40 \\ \hline
\end{tabular}
\end{center}
\label{tab3}
\end{table}
\vskip 0.1truecm

\begin{table}
\caption{Typical solutions for the fourth-generation mixing $(s_w,s_v,s_u)=
(0.3\lambda^3,0.3\lambda^2,\lambda)$ in case of $(m_{t'},m_{b'})=
(1200, 1170)$GeV for the three phases of $(\phi_1, \phi_2, \phi_3)$. Predictions 
of $B(K^+\to \pi^+\nu\bar{\nu})(10^{-10}), \Delta m_D({\rm in}  10^{-12}
{\rm MeV}), B(K_L\to \pi^0\nu\bar{\nu})(10^{-10})$ and the CP 
asymmetry for $B_d\to J/\psi K_S$ are added.}
\begin{center}
\begin{tabular}{c|c|c|c|c|c|c} \hline\hline
$\phi_1$ & $\phi_2$ & $\phi_3$ & $K^+\to \pi^+\nu\bar{\nu}$ & 
$\Delta m_D$ & $K_L\to \pi^0\nu\bar{\nu}$ & $C_f(B_d\to 
J/\psi K_S)$ \\ 
\hline
$5\pi/12$ & $\pi/6$ & $23\pi/12$ & 1.0 & 0.7 & 3.8 & 0.30 \\
$\pi/2$ & $\pi/3$ & 0 & 1.5 & 0.5 & 6.1 & 0.27 \\
$7\pi/12$ & $\pi/2$ & $\pi/6$ & 1.4 & 0.3 & 5.7 & 0.18 \\
$2\pi/3$ & $5\pi/6$ & $7\pi/12$ & 0.9 & 0.1 & 3.7 & 0.18 \\
$\pi/2$ & $\pi$ & $3\pi/4$ & 1.0 & 0.01 & 3.8 & 0.29 \\
$5\pi/12$ & $4\pi/3$ & $5\pi/3$ & 2.9 & 0.1 & 12.0 & 0.39 \\
$5\pi/12$ & $3\pi/2$ & $7\pi/4$ & 2.1 & 0.3 & 8.5 & 0.36 \\
$5\pi/12$ & $5\pi/3$ & $23\pi/12$ & 2.3 & 0.4 & 9.0 & 0.30 \\
$\pi/4$ & $11\pi/6$ & $\pi/6$ & 3.1 & 0.4 & 12.8 & 0.14 \\
$\pi/2$ & $2\pi$ & $7\pi/4$ & 1.1 & 0.5 & 4.2 & 0.35 \\ \hline
\end{tabular}
\end{center}
\label{tab4}
\end{table}
\vskip 0.1truecm

We show several typical solutions with respect to the three phases 
$(\phi_1, \phi_2, \phi_3)$ for the maximum mixing $(s_w,s_v,s_u)=
(0.8\lambda^3,0.8\lambda^2,\lambda)$ in the case of $(m_{t'},m_{b'})=
(400, 370)$GeV in Table 2, the ones for $(s_w,s_v,s_u)=
 (0.5\lambda^3,0.5\lambda^2,\lambda)$ in the case of $(m_{t'},m_{b'})=
(800, 770)$GeV in Table 3 and the ones for $(s_w,s_v,s_u)=
 (0.3\lambda^3,0.3\lambda^2,\lambda)$ in the case of $(m_{t'},m_{b'})=
(1200, 1170)$GeV in Table 4. The values of $(\phi_1, \phi_2, \phi_3)$ allowed 
by the constraints constitute a certain region in the plane, surrounding each of 
the solutions in Tables 2, 3 and 4. In the Tables, we also give the predictions 
of $B(K^+\to \pi^+\nu\bar{\nu}), \Delta m_D, B(K_L\to \pi^0
\nu\bar{\nu})$, and the CP asymmetry for $B_d\to J/\psi K_S$, 
which is explained in the following section, for each of the solutions. 

As can be seen from Tables 2, 3 and 4 for the "maximum" mixing of the fourth 
generation, the constraints from all the seven quantities considered here except 
$B(K^+\to \pi^+\nu\bar{\nu})$ could predict the values $(0.6-5.2)\times 
10^{-10}$ 
for $B(K^+\to \pi^+\nu\bar{\nu})$, including the values just outside 
the predictions of the Standard Model, $(0.6-1.5)\times 10^{-10}$, and 
not so large as the upper part of the measured value of $(0.7-13.9)\times 
10^{-10}$. This means that all the seven quantities except the present 
measurement of $B(K^+\to \pi^+\nu\bar{\nu})$ have already implied 
the fourth generation with the mixing as large as $(s_w,s_v,s_u) 
=(0.8\lambda^3,0.8\lambda^2,\lambda)$ for $m_{t'}=400$GeV and so on 
and that they could predict the quantities of $x_s, \Delta m_D$ and 
$B(K_L\to \pi^0\nu\bar{\nu})$ in the range of values shown in Table 5, 
which is explained in detail in the next section.
\vskip 0.7truecm

\centerline{\large\bf V  Discussions and conclusions}
\vskip 0.2truecm

\begin{table}
\caption{Comparison of $B(K^+\to\pi^+\nu\bar{\nu}), x_s(B_s-\bar{B_s} 
{\rm mixing}), \Delta m_D$ and $B(K_L\to\pi^0\nu\bar{\nu})$ among  
the experimental values, Standard Model(SM) predictions and four-generation 
model predictions.}
\begin{center}
\begin{tabular}{l|c|c|c|c}  \hline\hline
& $B(K^+\to\pi^+\nu\bar{\nu})$ & $x_s$ & $\Delta m_D({\rm MeV})$ 
& $B(K_L\to\pi^0\nu\bar{\nu})$  \\  \hline
Experiments & $\left( 4.2^{+9.7}_{-3.5}\right) \times 10^{-10}$ & $>12.8$ 
& $<1.4\times 10^{-10}$ & $<5.8\times 10^{-5}$  \\
SM & $(0.6-1.5)\times 10^{-10}$ & $19-27$ & $\sim 10^{-14}$ 
& $(1.1-5.0)\times 10^{-11}$  \\
4-generation & $(0.6-5.2)\times 10^{-10}$ & $19-29$ & $(0.01-2.1)
\times 10^{-12}$ & $(0.05-22)\times 10^{-10}$  \\  \hline
\end{tabular}
\end{center}
\label{tab5}
\end{table}
\vskip 0.1truecm

We can obtain the following predictions from these maximum mixings; the 
branching ratio of $K^+\to\pi^+\nu\bar{\nu}$ takes a range from the 
Standard Model(SM) values to the central value of the new measurement as $B=
(0.6-5.2)\times 10^{-10}$, the strength of $B_s-\bar{B_s}$ mixing 
is $19\le x_s\le 29$, where $x_s\equiv \Delta m_{B_s}/\Gamma_{B_s}$, 
$\Gamma_{B_s}$ being the total decay rate of $B_s$ meson, $\Delta m_D$ 
of $D^0-\bar{D^0}$ mixing could have a value $(0.01-2.1)\times 
10^{-12}$ MeV, extending to about two orders of magnitude 
larger than the SM prediction ($\sim 10^{-14}$ MeV\cite{Datta}), 
and the branching ratio of $K_L\to\pi^0\nu\bar{\nu}$ takes a range 
of $(0.05-22)\times 10^{-10}$, ranging 
from  the SM values to the values of two orders of magnitude larger than the SM 
prediction ($(1.1-5.0)\times 10^{-11}$\cite{Buchalla}). These results are 
summarized in Table 5.

The maximum mixing gives an interesting effect on the CP-asymmetry of the 
decay rates of the "gold-plate" mode of $B_d$ meson, $B_d\to J/\psi K_S$. 
The asymmetry is given by 
\begin{equation}
C_f=\frac{\Gamma (B_d\to J/\psi K_S)-\Gamma (\bar{B_d}\to J/\psi K_S)}
{\Gamma (B_d\to J/\psi K_S)+\Gamma (\bar{B_d}\to J/\psi K_S)},     
\label{shiki25}
\end{equation}
and it is expressed as\cite{Carter}
\begin{eqnarray}
C_f&=&-\frac{x_d}{1+x_d^2}{\rm Im}\Lambda=\frac{x_d}{1+x_d^2}
\sin 2\beta,    \label{shiki26}  \\
\Lambda&\equiv&{\sqrt \frac{M_{12}^*}{M_{12}}}
\frac{A(\bar{B_d}\to J/\psi K_S)}{A(B_d\to J/\psi K_S)},  \label{shiki27}
\end{eqnarray}
where $x_d$ is the mixing strength of $B_d-\bar{B_d}$ mixing, $M_{12}$ 
the off-diagonal element of the mass matrix in $B_d-\bar{B_d}$ system, 
$A$ the decay amplitude and $\beta$ is one of the angles of the unitarity triangle. 
In the Standard Model\cite{Dunietz}, the quantity $C_f$ takes a positive sign as 
$0.18\le C_f\le 0.37$ for $B_d\to J/\psi K_S$, resulting 
from the phase range of $0<\phi_1<\pi$, which is constrained from the positive 
sign of the CP-violating parameter $\varepsilon_K$. 
However, in the four-generation model\cite{Hasuike}, $C_f$ can take also 
a negative sign as $-0.38\le C_f\le 0.40$, since the phase $\phi_1$ takes 
the whole range of $0<\phi_1<2\pi$ due to the occurence of the two more 
new phases $\phi_2$ and $\phi_3$. For the moment, $\sin 2\beta$ of 
eq.(\ref{shiki26}) has recently been measured to be positive as 
$\sin 2\beta=\left( 3.2^{+1.8}_{-2.0}\pm 0.5 \right)$ 
by OPAL Collaboration\cite{Akerstaff} and $\sin 2\beta=
(1.8\pm 1.1\pm 0.3)$ by CDF Collaboration\cite{Abe}, which means that 
$C_f$ is positive as in the Standard Model. We should add that 
although the penguin diagrams could affect 
the decay amplitude in the four-generation model, they would bring at most 
several percent change of $C_f$. 

The unitarity triangle in the Standard Model transforms into unitarity 
quadrangle in the four-generation model\cite{Nir}. For the "maximum" mixing 
obtained here, some of the typical quadrangles are shown in Fig.4 for $m_{t'}=
400$GeV and in Fig.5 for $m_{t'}=800$GeV. The fourth 
side of the quadrangle, $V_{t'd}V_{t'b}^*$, is of order $\lambda^4$, while 
the other three sides are of order $\lambda^3$. The first two quadrangles of 
Figs.4 and 5 are for positive sign of $C_f$. The third ones of Figs.4 and 5 
are for negative sign of $C_f$ and are reversed with respect to the base 
line $V_{cd}V_{cb}^*$, since $\phi_1>\pi$, where $\phi_1$ 
corresponds to the anti-clockwise angle measured from $V_{cd}V_{cb}^*$ 
to $V_{ud}V_{ub}^*$ and $\phi_3$ to the anti-clockwise angle from 
$V_{cd}V_{cb}^*$ to $V_{t'd}V_{t'b}^*$. Incidentally, the quadrangles 
for the solutions with smaller mixing of $(s_w,s_v,s_u)= (\lambda^4,
\lambda^3,\lambda^2)$ for $m_{t'}=400$GeV are given in Fig.6. In this 
case, the size of the fourth side, $V_{t'd}V_{t'b}^*$, is of order $\lambda^6$ 
and is about 1/100 that of the side $V_{cd}V_{cb}^*$ and the quadrangle 
could not be distinguished from the triangle, and the branching ratio of 
$K^+\to\pi^+\nu\bar{\nu}$ is predicted to be in the range of $(0.6-1.2)
\times 10^{-10}$, which agrees with the predictions of the Standard Model. 
So, if the future measurements of $K^+\to\pi^+\nu\bar{\nu}$ show its 
branching ratio to be in the range of the Standard Model values, the large mixing 
of the fourth generation obtained here as the "maximum" one will not be allowed. 

We should remark that this large mixing of the fourth generation we found here 
is not due to the fairly large theoretical uncertainties in $B_K=0.75\pm 0.15$ 
and $f_B\sqrt{B_B}=(0.20\pm 0.04)$GeV. Even if we prescribe to reduce the 
uncertainties of these quantities by $1/3$ to $1/4$ keeping the central values as 
$B_K=0.75\pm 0.05$ and $f_B\sqrt{B_B}=(0.20\pm 0.01)$GeV, we can 
still find some of the solutions such as listed in Tables 6 and 7 for $m_{t'}=
400$GeV and 800GeV, respectively. 

\begin{table}
%\caption{The same as in Table 2 except that $B_K=0.75\pm 0.05$ and 
%$\bar{\nu}$ and $\sqrt B$ and $(0.20\pm 0.01)$.}
\caption{The same as in Table 2 except that $B_K=0.75\pm 0.05$ and $f_B^2
B_B = (0.20\pm 0.01)^2 {\rm GeV}^2$.}
%\caption{Some solutions for the fourth-generation mixing $(s_w,s_v,s_u)=
%(0.8\lambda^3,0.8\lambda^2,\lambda)$ in case of $(m_{t'},m_{b'})=
%(400, 350)$GeV for $B_K=0.75\pm 0.05$ and $f_B\sqrt{B_B}=
%(0.20\pm 0.01)$GeV.}
\begin{center}
\begin{tabular}{c|c|c|c|c|c|c} \hline\hline
$\phi_1$ & $\phi_2$ & $\phi_3$ & $K^+\to \pi^+\nu\bar{\nu}$ & 
$\Delta m_D$ & $K_L\to \pi^0\nu\bar{\nu}$ & $C_f(B_d\to 
J/\psi K_S)$ \\ 
\hline
$\pi/4$ & $\pi/3$ & 0 & 2.2 & 1.8 & 8.7 & 0.24 \\
$\pi/2$ & $5\pi/6$ & $\pi/2$ & 1.7 & 0.7 & 7.0 & 0.25 \\
$\pi/2$ & $7\pi/6$ & $11\pi/12$ & 1.7 & 0.4 & 6.6 & 0.30 \\  \hline
\end{tabular}
\end{center}
\label{tab6}
\end{table}
\vskip 0.1truecm

\begin{table}
\caption{The same as in Table 3 except that $B_K=0.75\pm 0.05$ and $f_B^2
B_B = (0.20\pm 0.01)^2 {\rm GeV}^2$.}
%\caption{Some solutions for the fourth-generation mixing $(s_w,s_v,s_u)=
%(0.5\lambda^3,0.5\lambda^2,\lambda)$ in case of $(m_{t'},m_{b'})=
%(800, 750)$GeV for $B_K=0.75\pm 0.05$ and $f_B\sqrt{B_B}=
%(0.20\pm 0.01)$GeV.}
\begin{center}
\begin{tabular}{c|c|c|c|c|c|c} \hline\hline
$\phi_1$ & $\phi_2$ & $\phi_3$ & $K^+\to \pi^+\nu\bar{\nu}$ & 
$\Delta m_D$ & $K_L\to \pi^0\nu\bar{\nu}$ & $C_f(B_d\to 
J/\psi K_S)$ \\ 
\hline
$\pi/2$ & $5\pi/6$ & $\pi/2$ & 2.7 & 0.3 & 11.0 & 0.22 \\
$\pi/2$ & $\pi$ & $4\pi/3$ & 4.8 & 0.1 & 19.8 &0.38 \\
$15\pi/8$ & $3\pi/2$ & 0 & 5.1 & 0.4 & 19.5 & -0.13 \\ \hline
\end{tabular}
\end{center}
\label{tab7}
\end{table}
\vskip 0.1truecm

Summarizing, we find "maximum" mixings of the fourth generation 
$(V_{t'd},V_{t's},V_{t'b})\simeq (0.8\lambda^3, 0.8\lambda^2, \lambda)$ 
for $(m_{t'},m_{b'})=(400, 370)$GeV, $(V_{t'd},V_{t's},V_{t'b})
\simeq (0.5\lambda^3, 0.5\lambda^2, \lambda)$ for $(m_{t'},m_{b'})=
(800, 770)$GeV and $(V_{t'd},V_{t's},V_{t'b})\simeq (0.3\lambda^3, 
0.3\lambda^2,  \lambda)$ 
for $(m_{t'},m_{b'})=(1200, 1170)$GeV, which are 
consistent with the eight constraints of $\Delta m_K, \varepsilon_K, 
B_d-\bar{B_d}$ mixing, $K^+\to\pi^+\nu\bar{\nu}, B_s-\bar{B_s}$ 
mixing, $D^0-\bar{D^0}$ mixing, $K_L\to\pi^0\nu\bar{\nu}$ and 
$(K_L\to\mu\bar{\mu})_{\rm SD}$. The mass difference 
$\Delta m_D$ from $D^0-\bar{D^0}$ mixing and the branching ratio of
$K_L\to\pi^0\nu\bar{\nu}$ could reach the values one to two orders of 
magnitude larger than the Standard Model predictions, and the CP asymmetry 
of the decay rates of $B_d\to J/\psi K_S$ could take a value of opposite 
sign to the SM one. Measurements of $\Delta m_D$ and $B(K_L\to\pi^0
\nu\bar{\nu})$ are expected to be done and further data of $B(K^+\to
\pi^+\nu\bar{\nu})$ with more statistics are required. 

We are grateful to Takeshi Komatsubara, Minoru Tanaka, Takeshi Kurimoto, 
Xing Zhi-Zhong, Masako Bando, C.S. Lim, and Morimitsu Tanimoto for 
helpful discussions.

\newpage
\centerline{\bf Figure captions}

\vskip 1.0truecm
\noindent
{\bf Fig.1.} $W-W$ box diagram for $K_L-K_S$ mass difference in the 
four-generation model.
\vskip 0.5truecm

\noindent
{\bf Fig.2.} $W-W$ box and $Z^0$-penguin diagrams for $K^+\to\pi^+
\nu\bar{\nu}$.
\vskip 0.5truecm

\noindent
{\bf Fig.3.} The dominant $W-W$ box diagram for $D^0-\bar{D^0}$ 
mixing in the four-generation model.
\vskip 0.5truecm

\noindent
{\bf Fig.4.} Typical examples of the unitarity quadrangle for $(s_w,s_v,s_u)=
(0.8\lambda^3,0.8\lambda^2,\lambda)$ in case of $(m_{t'},m_{b'})=
(400, 370)$GeV. (a) $\phi_1=\frac{\pi}{2}, \phi_2=2\pi, \phi_3=
\frac{7}{4}\pi; B(K^+\to\pi^+\nu\bar{\nu})=1.6\times 10^{-10}, 
C_f(B_d\to J/\psi K_S)=0.34$, (b) $\phi_1=\frac{\pi}{2}, \phi_2=
\frac{5}{6}\pi, \phi_3=\frac{\pi}{2}; B(K^+\to\pi^+\nu\bar{\nu})=1.7
\times 10^{-10}, C_f(B_d\to J/\psi K_S)=0.25$,  (c) $\phi_1=\frac{13}{8}
\pi, \phi_2=\pi, \phi_3=\frac{3}{4}\pi; B(K^+\to\pi^+\nu\bar{\nu})=2.7
\times 10^{-10}, C_f(B_d\to J/\psi K_S)=-0.35$.
\vskip 0.5truecm

\noindent
{\bf Fig.5.} Typical examples of the unitarity quadrangle for $(s_w,s_v,s_u)=
(0.5\lambda^3,0.5\lambda^2,\lambda)$ in case of $(m_{t'},m_{b'})=
(800, 770)$GeV. (a) $\phi_1=\frac{\pi}{2}, \phi_2=\frac{3}{2}\pi, 
\phi_3=\frac{11}{6}\pi; B(K^+\to\pi^+\nu\bar{\nu})=4.6\times 10^{-10}, 
C_f(B_d\to J/\psi K_S)=0.32$, (b) $\phi_1=\frac{5}{12}\pi, \phi_2=
\frac{2}{3}\pi, \phi_3=\frac{\pi}{3}; B(K^+\to\pi^+\nu\bar{\nu})=2.7
\times 10^{-10}, C_f(B_d\to J/\psi K_S)=0.17$,  (c) $\phi_1=\frac{19}{12}
\pi, \phi_2=\frac{11}{6}\pi, \phi_3=\frac{\pi}{3}\pi; B(K^+\to\pi^+
\nu\bar{\nu})=5.2\times 10^{-10}, C_f(B_d\to J/\psi K_S)=-0.40$.
\vskip 0.5truecm

\noindent
{\bf Fig.6.} Typical examples of the unitarity quadrangle for $(s_w,s_v,s_u)=
(\lambda^4,\lambda^3,\lambda^2)$ in case of $(m_{t'},m_{b'})=
(400, 370)$GeV. (a) $\phi_1=\frac{\pi}{2}, \phi_2=\frac{\pi}{6}, \phi_3=
\frac{7}{4}\pi; B(K^+\to\pi^+\nu\bar{\nu})=0.94\times 10^{-10}, 
C_f(B_d\to J/\psi K_S)=0.30$, (b) $\phi_1=\frac{\pi}{4}, \phi_2=
\frac{\pi}{6}\pi, \phi_3=\frac{3}{4}\pi; B(K^+\to\pi^+\nu\bar{\nu})=
0.89\times 10^{-10}, C_f(B_d\to J/\psi K_S)=0.28$,  (c) $\phi_1=\frac{\pi}
{4}, \phi_2=\frac{\pi}{3}, \phi_3=\frac{5}{4}\pi; B(K^+\to\pi^+\nu
\bar{\nu})=1.0\times 10^{-10}, C_f(B_d\to J/\psi K_S)=0.29$.
\vskip 0.5truecm

\newpage

\begin{figure}
\begin{center}
{\unitlength=1mm
\begin{picture}(100,50)
\thicklines
\put(14,12){\line(1,0){6}}
\put(35,12){\vector(-1,0){15}}
\put(14,37){\vector(1,0){10}}
\put(24,37){\line(1,0){11}}
\put(35,12){\line(0,1){11}}
\put(35,37){\vector(0,-1){14}}
\multiput(35,12)(5.2,0){5}{\line(2,-3){2.6}}
\multiput(37.6,8.1)(5.2,0){5}{\line(2,3){2.6}}
\multiput(35,37)(5.2,0){5}{\line(2,3){2.6}}
\multiput(37.6,40.9)(5.2,0){5}{\line(2,-3){2.6}}
\put(61.0,12){\line(1,0){12}}
\put(82.0,12){\vector(-1,0){9}}
\put(61.0,37){\vector(1,0){15}}
\put(76.0,37){\line(1,0){6}}
\put(61.0,12){\vector(0,1){14}}
\put(61.0,26){\line(0,1){11}}
\put(9,12){\makebox(0,0){\large $\bar{\rm d}$}}
\put(9,37){\makebox(0,0){\large s}}
\put(87.0,12){\makebox(0,0){\large $\bar{\rm s}$}}
\put(87.0,37){\makebox(0,0){\large d}}
\put(48.0,2.1){\makebox(0,0){\large W}}
\put(48.0,46.9){\makebox(0,0){\large W}}
\put(26,24){\makebox(0,0){\large c,t,t'}}
\put(70.0,24){\makebox(0,0){\large c,t,t'}}
\end{picture}}
\end{center}
\caption{ }
\label{Fig1}
\end{figure}

\newpage

\begin{figure}
\begin{center}
{\unitlength=1mm
\begin{picture}(150,150)
\thicklines
\put(9,56){\makebox(0,0){\large u}}
\put(12,56){\vector(1,0){27}}
\put(39,56){\line(1,0){25}}
\put(67,56){\makebox(0,0){\large u}}
\put(9,44){\makebox(0,0){\large $\bar{\rm s}$}}
\put(12,44){\line(1,0){5}}
\put(25,44){\vector(-1,0){8}}
\multiput(25,44)(5.2,0){5}{\line(2,3){2.6}}
\multiput(27.6,47.9)(5.2,0){5}{\line(2,-3){2.6}}
\put(51,44){\line(1,0){7}}
\put(64,44){\vector(-1,0){6}}
\put(67,44){\makebox(0,0){\large $\bar{\rm d}$}}
\put(38,51){\makebox(0,0){\large W}}
\put(25,44){\line(1,-1){6}}
\put(38,31){\vector(-1,1){7}}
\put(38,31){\line(1,1){6}}
\put(51,44){\vector(-1,-1){7}}
\put(22,37){\makebox(0,0){\large $\bar{\rm c},\bar{\rm t},\bar{\rm t'}$}}
\put(54,37){\makebox(0,0){\large $\bar{\rm c},\bar{\rm t},\bar{\rm t'}$}}
\multiput(38,31)(0,-4.8){3}{\line(3,-2){3.6}}
\multiput(38,26.2)(0,-4.8){3}{\line(3,2){3.6}}
\put(38,16.6){\vector(-2,-3){4.4}}
\put(33.6,10){\line(-2,-3){2.4}}
\put(38,16.6){\line(2,-3){2.4}}
\put(44.8,6.4){\vector(-2,3){4.4}}
\put(34,23.8){\makebox(0,0){\large Z}}
\put(29.2,3.4){\makebox(0,0){\large $\nu $}}
\put(46.8,3.4){\makebox(0,0){\large $\bar{\nu}$}}
\put(9,124){\makebox(0,0){\large u}}
\put(12,124){\vector(1,0){27}}
\put(39,124){\line(1,0){25}}
\put(67,124){\makebox(0,0){\large u}}
\put(9,112){\makebox(0,0){\large $\bar{\rm s}$}}
\put(12,112){\line(1,0){5}}
\put(37,112){\vector(-1,0){20}}
\put(58,112){\vector(-1,0){21}}
\put(64,112){\vector(-1,0){6}}
\put(67,112){\makebox(0,0){\large $\bar{\rm d}$}}
\multiput(25,112)(0,-4.8){5}{\line(3,-2){3.6}}
\multiput(25,107.2)(0,-4.8){5}{\line(3,2){3.6}}
\multiput(51,112)(0,-4.8){5}{\line(3,-2){3.6}}
\multiput(51,107.2)(0,-4.8){5}{\line(3,2){3.6}}
\put(25,88){\line(1,0){12}}
\put(51,88){\vector(-1,0){14}}
\put(25,88){\vector(1,-1){7}}
\put(32,81){\line(1,-1){5}}
\put(51,88){\line(1,-1){5}}
\put(63,76){\vector(-1,1){7}}
\put(41,117){\makebox(0,0){\large $\bar{\rm c},\bar{\rm t},\bar{\rm t'}$}}
\put(20,99.8){\makebox(0,0){\large W}}
\put(58.7,99.8){\makebox(0,0){\large W}}
\put(41,92.6){\makebox(0,0){\large ${\rm e},\mu,\tau $}}
\put(40,73){\makebox(0,0){\large $\nu $}}
\put(66,73){\makebox(0,0){\large $\bar{\nu}$}}
\put(79,124){\makebox(0,0){\large u}}
\put(82,124){\vector(1,0){27}}
\put(109,124){\line(1,0){25}}
\put(137,124){\makebox(0,0){\large u}}
\put(79,112){\makebox(0,0){\large $\bar{\rm s}$}}
\put(82,112){\line(1,0){5}}
\put(107,112){\vector(-1,0){20}}
\put(128,112){\vector(-1,0){21}}
\put(134,112){\vector(-1,0){6}}
\put(137,112){\makebox(0,0){\large $\bar{\rm d}$}}
\multiput(95,112)(2.88,-4.32){4}{\line(5,-1){3.6}}
\multiput(97.88,107.68)(2.88,-4.32){3}{\line(1,5){0.72}}
\multiput(119.48,112)(-2.88,-4.32){4}{\line(-5,-1){3.6}}
\multiput(116.6,107.68)(-2.88,-4.32){3}{\line(-1,5){0.72}}
\multiput(107.24,98.32)(0,-4.8){3}{\line(3,-2){3.6}}
\multiput(107.24,93.52)(0,-4.8){3}{\line(3,2){3.6}}
\put(107.24,83.92){\vector(-2,-3){4.4}}
\put(102.84,77.32){\line(-2,-3){2.4}}
\put(107.24,83.92){\line(2,-3){2.4}}
\put(114.04,73.72){\vector(-2,3){4.4}}
\put(108,117){\makebox(0,0){\large $\bar{\rm c},\bar{\rm t},\bar{\rm t'}$}}
\put(96,104){\makebox(0,0){\large W}}
\put(119,104){\makebox(0,0){\large W}}
\put(104,91.8){\makebox(0,0){\large Z}}
\put(98.44,70.72){\makebox(0,0){\large $\nu $}}
\put(116.04,70.72){\makebox(0,0){\large $\bar{\nu}$}}
\end{picture}}
\end{center}
\caption{ }
\label{Fig2}
\end{figure}

\newpage

\begin{figure}
\begin{center}
{\unitlength=1mm
\begin{picture}(100,50)
\thicklines
\put(14,12){\line(1,0){6}}
\put(35,12){\vector(-1,0){15}}
\put(14,37){\vector(1,0){10}}
\put(24,37){\line(1,0){11}}
\put(35,12){\line(0,1){11}}
\put(35,37){\vector(0,-1){14}}
\multiput(35,12)(5.2,0){5}{\line(2,-3){2.6}}
\multiput(37.6,8.1)(5.2,0){5}{\line(2,3){2.6}}
\multiput(35,37)(5.2,0){5}{\line(2,3){2.6}}
\multiput(37.6,40.9)(5.2,0){5}{\line(2,-3){2.6}}
\put(61.0,12){\line(1,0){12}}
\put(82.0,12){\vector(-1,0){9}}
\put(61.0,37){\vector(1,0){15}}
\put(76.0,37){\line(1,0){6}}
\put(61.0,12){\vector(0,1){14}}
\put(61.0,26){\line(0,1){11}}
\put(9,12){\makebox(0,0){\large $\bar{\rm u}$}}
\put(9,37){\makebox(0,0){\large c}}
\put(87.0,12){\makebox(0,0){\large $\bar{\rm c}$}}
\put(87.0,37){\makebox(0,0){\large u}}
\put(48.0,2.1){\makebox(0,0){\large W}}
\put(48.0,46.9){\makebox(0,0){\large W}}
\put(29,24){\makebox(0,0){\large b'}}
\put(67,24){\makebox(0,0){\large b'}}
\end{picture}}
\end{center}
\caption{ }
\label{Fig3}
\end{figure}

\newpage

\begin{figure}
\begin{center}
{\unitlength=1mm
\begin{picture}(120,200)
\thicklines
\put(90,40){\vector(-1,0){70}}
\put(20,40){\line(2,-5){9.52}}
\put(29.52,16.2){\line(6,1){72.24}}
\put(101.76,28.24){\line(-1,1){11.76}}
\put(13,25){\makebox(0,0){\large $V_{ud}V_{ub}^*$}}
\put(65,15){\makebox(0,0){\large $V_{td}V_{tb}^*$}}
\put(108,35){\makebox(0,0){\large $V_{t'd}V_{t'b}^*$}}
\put(55,46){\makebox(0,0){\large $V_{cd}V_{cb}^*$}}
\put(15,58){\makebox(0,0){\large (c)}}
\put(90,90){\vector(-1,0){70}}
\put(20,90){\line(0,1){26}}
\put(20,116){\line(5,-3){70}}
\put(90,74){\line(0,1){16}}
\put(10,102){\makebox(0,0){\large $V_{ud}V_{ub}^*$}}
\put(49,107){\makebox(0,0){\large $V_{td}V_{tb}^*$}}
\put(101,81){\makebox(0,0){\large $V_{t'd}V_{t'b}^*$}}
\put(50,84){\makebox(0,0){\large $V_{cd}V_{cb}^*$}}
\put(15,129){\makebox(0,0){\large (b)}}
\put(90,150){\vector(-1,0){70}}
\put(20,150){\line(0,1){25}}
\put(20,175){\line(4,-1){60}}
\put(80,160){\line(1,-1){10}}
\put(10,162){\makebox(0,0){\large $V_{ud}V_{ub}^*$}}
\put(50,174){\makebox(0,0){\large $V_{td}V_{tb}^*$}}
\put(95,157){\makebox(0,0){\large $V_{t'd}V_{t'b}^*$}}
\put(55,144){\makebox(0,0){\large $V_{cd}V_{cb}^*$}}
\put(15,185){\makebox(0,0){\large (a)}}
\end{picture}}
\end{center}
\caption{ }
\label{Fig4}
\end{figure}

\newpage

\begin{figure}
\begin{center}
{\unitlength=1mm
\begin{picture}(120,200)
\thicklines
\put(90,50){\vector(-1,0){70}}
\put(20,50){\line(1,-4){6}}
\put(26,26){\line(4,1){58.35}}
\put(84.35,40.59){\line(3,5){5.65}}
\put(13,35){\makebox(0,0){\large $V_{ud}V_{ub}^*$}}
\put(55,27){\makebox(0,0){\large $V_{td}V_{tb}^*$}}
\put(97,45){\makebox(0,0){\large $V_{t'd}V_{t'b}^*$}}
\put(55,56){\makebox(0,0){\large $V_{cd}V_{cb}^*$}}
\put(15,65){\makebox(0,0){\large (c)}}
\put(90,90){\vector(-1,0){70}}
\put(20,90){\line(1,4){6}}
\put(26,114){\line(2,-1){60.31}}
\put(86.31,83.85){\line(3,5){3.69}}
\put(13,102){\makebox(0,0){\large $V_{ud}V_{ub}^*$}}
\put(55,107){\makebox(0,0){\large $V_{td}V_{tb}^*$}}
\put(97,86){\makebox(0,0){\large $V_{t'd}V_{t'b}^*$}}
\put(50,84){\makebox(0,0){\large $V_{cd}V_{cb}^*$}}
\put(15,129){\makebox(0,0){\large (b)}}
\put(90,150){\vector(-1,0){70}}
\put(20,150){\line(0,1){25.47}}
\put(20,175.47){\line(3,-1){62}}
\put(82,154.8){\line(5,-3){8}}
\put(10,162){\makebox(0,0){\large $V_{ud}V_{ub}^*$}}
\put(52,171){\makebox(0,0){\large $V_{td}V_{tb}^*$}}
\put(95,155){\makebox(0,0){\large $V_{t'd}V_{t'b}^*$}}
\put(55,144){\makebox(0,0){\large $V_{cd}V_{cb}^*$}}
\put(15,185){\makebox(0,0){\large (a)}}
\end{picture}}
\end{center}
\caption{ }
\label{Fig5}
\end{figure}

\newpage

\begin{figure}
\begin{center}
{\unitlength=1mm
\begin{picture}(120,200)
\thicklines
\put(90,30){\vector(-1,0){70}}
\put(20,30){\line(1,1){21.1}}
\put(41.1,51.1){\line(3,-1){52.5}}
\put(90,30){\line(1,1){3.6}}
\put(23,43){\makebox(0,0){\large $V_{ud}V_{ub}^*$}}
\put(69,48){\makebox(0,0){\large $V_{td}V_{tb}^*$}}
\put(102,30){\makebox(0,0){\large $V_{t'd}V_{t'b}^*$}}
\put(55,24){\makebox(0,0){\large $V_{cd}V_{cb}^*$}}
\put(15,65){\makebox(0,0){\large (c)}}
\put(90,90){\vector(-1,0){70}}
\put(20,90){\line(1,1){18.44}}
\put(38.44,108.44){\line(5,-2){55.2}}
\put(90,90){\line(1,-1){3.64}}
\put(22,102){\makebox(0,0){\large $V_{ud}V_{ub}^*$}}
\put(67,104){\makebox(0,0){\large $V_{td}V_{tb}^*$}}
\put(102,89){\makebox(0,0){\large $V_{t'd}V_{t'b}^*$}}
\put(55,84){\makebox(0,0){\large $V_{cd}V_{cb}^*$}}
\put(15,125){\makebox(0,0){\large (b)}}
\put(90,150){\vector(-1,0){70}}
\put(20,150){\line(0,1){25.8}}
\put(20,175.8){\line(3,-1){66.3}}
\put(90,150){\line(-1,1){3.7}}
\put(10,162){\makebox(0,0){\large $V_{ud}V_{ub}^*$}}
\put(56,170){\makebox(0,0){\large $V_{td}V_{tb}^*$}}
\put(99,152){\makebox(0,0){\large $V_{t'd}V_{t'b}^*$}}
\put(55,144){\makebox(0,0){\large $V_{cd}V_{cb}^*$}}
\put(15,185){\makebox(0,0){\large (a)}}
\end{picture}}
\end{center}
\caption{ }
\label{Fig6}
\end{figure}

\end{document}